\documentclass[aps,prb,12pt]{revtex4}
\usepackage{graphicx}
\usepackage{amsmath}
\usepackage{amsfonts}
\usepackage{amssymb}
\begin{document}
\title{Quantum Mechanics and Common Sense}
\author{S.V.Gantsevich}
\affiliation{Ioffe Institute, Russian Academy of Sciences, 194021
Saint Petersburg, Russia, e-mail: sergei.elur@mail.ioffe.ru}
\begin{abstract}
\baselineskip=2.5ex {\it A physical picture for Quantum Mechanics
which permits to conciliate it with the usual common sense is
proposed. The picture agrees with the canonical Copenhagen
interpretation making more clear its statements.}
\end{abstract} \vspace{0.7truecm}
\maketitle \baselineskip=2.5ex Pacs:  03.65.-w, 03.65.Ud\\
\par
It is generally accepted that Quantum Mechanics is
"counterintuitive"\, or, simply speaking, it contradicts our
ordinary common sense based on everyday experience [\cite{Zel,Omn,
Zr,Prs,Schl}].
\par
 Is this situation a peculiar feature of Quantum Mechanics?
\par
The answer is NO.
\par
The science history tells us that this situation is rather a rule
than an exception. There were always puzzles and mysteries in
Science. But after some time (years, tens of years or even
centuries) they vanished or transformed into trivialities of no
mention.
\par {\it Such lot is inevitable also for Quantum
Mechanics.}
\par
It is important to notice that in all cases of former mysteries
their origin was always the {\it wrong picture of observed
phenomena.} And this wrong picture always seemed so {\it natural
and self-obvious} that any doubt of its validity never arouse.
When such invisible wall preventing the adequate understanding was
broken in some or another way all mysteries and puzzles vanished
and became fully forgotten.
\par
It is reasonable to conclude that such {\it wrong physical picture
exists in Quantum Mechanics} and just this is the cause of all
difficulties. Thus the question is what is wrong and how to repair
the situation. The answer is very simple but quite unexpected.
\par
The Founders of QM and their followers frequently used to say that
one should reject habitual concepts of the pre-quantum era and
rely entirely on the mathematics and logics
[\cite{B1,B2,BH,LB,Bm,Schr}]. However, at one point they did
exactly opposite and carefully preserved such concept though QM
formulae and logics were against it. Unfortunately it was just the
point which made impossible any simple and reasonable treatment of
quantum phenomena.
\par
What is this unfortunate point? Let us see.
\par
The world around us is classical. It consists of observed or
measured physical quantities. We have also strict quantum
mechanical rules for the calculations of these quantities. So far
there were no contradictions with these rules and experiments.
Therefore logically we should regard the QM formula for the
observable quantity as the center of our physical considerations.
For one quantum object this formula has the form:
\begin{equation}\label{1}
\bar{c}=\langle \psi|C|\psi\rangle\equiv\int_V d{\bf r}d{\bf
r'}\psi\dag({\bf r'},t)C({\bf r'}|{\bf r})\psi({\bf r},t)
\end{equation}
Here $C\equiv C({\bf r'}|{\bf r})$ is the operator of an
observable quantity. It determines the values of the physical
quantity in our world. Two other components are the wave function
$\psi\equiv |\psi\rangle$ and the complex conjugated wave function
$\psi^\dag\equiv\langle\psi|$. Thus it seems that the wave
function $\psi$ determines the state of the quantum object and we
should take it as the basis of our physical picture. This approach
to the interpretation of quantum phenomena looks quite natural and
self-obvious. All QM books and papers accept it. Nevertheless it
is wrong and leads to an impasse. The expression (\ref{1})
contains {\it two different quantum entities} $\psi$ and
$\psi^\dag$. And just these two entities determine the quantum
state and (together with the operator $C$) the physical quantities
that we observe in our world. A single $\psi$ as well as a single
$\psi^\dag$ are unobservable in our world. They are the elements
of Quantum World and appear in our Classical World only by pairs.
Thus a classical device is necessary for a measurement in QM as is
rightly stated in the Copenhagen interpretation. Without it we
shall see nothing.
\par
The peculiarity of Quantum Mechanics is the {\it linearity of
equations for the wave functions} and the {\it bi-linearity of
observable physical quantities}. This fact leads to the apparition
of two different types of physical states.
\par
Let us take, for simplicity, $C({\bf r'}|{\bf r})=U({\bf
r})\delta({\bf r-r'})$ and $U({\bf r})=\delta({\bf r-R})$. Then we
get from (\ref{1}):
\begin{equation}\label{2}
\overline{U}=\langle \psi|U|\psi\rangle\equiv\int_Vd{\bf
r}\psi^\dag({\bf r},t)U({\bf r})\psi({\bf r},t)=|\psi({\bf
R},t)|^2
\end{equation}
\par
It is convenient to call $\psi^\dag$ and $\psi$ as the bra and ket
functions or simply the {\it bra} and {\it ket}.
\par
Let us emphasize once more that the bra and ket in (\ref{1}) or
(\ref{2}) are independent quantities. They have generally
different space and time coordinates and their independent time
evolutions are governed by the separate equations of motion
($t\geq 0$):
\begin{eqnarray}\label{3}
(\partial_t+iH)\psi({\bf r},t)=\psi({\bf r})\delta(t)\\\nonumber
\psi({\bf r},t)=\frac{1}{\partial_t+iH}\psi({\bf
r})\delta(t)=e^{-iHt}\psi({\bf r})
\end{eqnarray}
The analogous formulae govern the evolution of the bra-function
$\psi^\dag({\bf r},t)$.
\par
The initial values $\psi({\bf r})$ and $\psi^\dag({\bf r'})$ can
be represented as the sums of the eigenfunctions of the
Hamiltonian $H\psi_p=\epsilon_p\psi_p$. The space forms of
$\psi^\dag_p$ and $\psi_p$ remain unchanged and only their phases
vary with time:
\begin{equation}\label{4}
\psi({\bf r},t)=\sum_p a_pe^{-i\epsilon_pt}\psi_p({\bf r})
\end{equation}
Substituting $\psi^\dag$ and $\psi$ into (\ref{2}) we get:
\begin{eqnarray}\label{5}
|\psi({\bf R},t)|^2=\sum_{p'p} a_{p'}^\dag a_p
e^{i[(\epsilon_{p'}-\epsilon_p)t]}\psi^\dag_{p'}({\bf
R})\psi_p({\bf R})=\\\nonumber =\sum_p |a_{p}|^2|\psi_p({\bf
R})|^2+\sum_{p'\neq p} a_{p'}^\dag a_p
e^{i[(\epsilon_{p'}-\epsilon_p)t]}\psi^\dag_{p'}({\bf
R})\psi_p({\bf R})
\end{eqnarray}
Suppose that the eigenfunctions of $H$ are orthogonal and
normalized. Then after the integration over all space only the
diagonal ($p=p'$) part of (\ref{5}) survives:
\begin{equation}\label{6}
\int_V|\psi({\bf R},t)|^2d{\bf R}=\sum_p |a_p|^2\equiv\sum_p F_p
\end{equation}
The quantity $F_p$ is the occupancy number of the state $p$. It
gives the probability to find a quantum particle in the given
state.
\par
It is important that the occupancy number $F_p$ is formed by {\it
two entities - one bra and one ket} with the same quantum indices.
It does not depend on the initial phase as well as on the phase
acquired during the time evolution:
\begin{equation}\label{7}
F_p(t)=a_p^\dag e^{i\epsilon_pt}a_pe^{-i\epsilon_pt}=|a_p|^2\equiv
F_p
\end{equation}
Now consider the non-diagonal elements ($p\neq p'$) of the sum
(\ref{5}), (for brevity ${\bf R}\equiv x$):
\begin{equation}\label{8}
\sum_{p'\neq p}\langle p'|U|p\rangle=\sum_{p'\neq p}
a^\dag_{p'}a_p e^{i[(\epsilon_{p'}-\epsilon_p)t} \psi^\dag_{p'}
(x)\psi_p(x)
\end{equation}
Unlike the expression (\ref{7}) all terms of this expression are
phase dependent. They vanish after the averaging over phases:
\begin{equation}\label{9}
\sum_{p'\neq p} |a^\dag_{p'}a_p| |\psi^\dag_{p'}
(x)\psi_p(x)|\overline{e^{i(\varphi_p-\varphi_{p'})}}=0
\end{equation}
The phase differences in (\ref{9}) may be generally regarded as
random quantities, so we come to the following probability rules
for the pure ($p=p'$) and mixed ($p\neq p'$) quantum states. For
pure states we have the Born rule:
\begin{equation}\label{10}
w_p(x)=|\psi_p(x)|^2F_p
\end{equation}
The contribution of one mixed state in (\ref{8}) is not real since
it is proportional to $exp(i\phi)$ where
$\phi=\varphi_p(x,t)-\varphi_{p'}(x,t)$ is the phase difference
between the bra and ket at time $t$ in the space point $x$. The
sum $\langle p|U|p'\rangle$ + $\langle p'|U|p\rangle$ is real so
we can use it to generalize the Born rule for mixed quantum states
splitting $\cos\phi$ into two positive parts:
\begin{equation}\label{11}
\cos\phi=\cos^2(\phi/2)-\sin^2(\phi/2)=P-Q
\end{equation}
We can regard $P$ as the probability of positive result and $Q$ as
the probability of negative result with $P+Q=1$.
\par
The expressions (\ref{10}) and (\ref{11}) show that pure and mixed
states play different roles when their bra and ket meet and appear
in our world. The "pure"\, (bra+ket) pairs are phase independent
and always give the same result. The "mixed"\, pairs are phases
dependent and give results of different signs. For arbitrary phase
difference their contributions vanish in average:
\begin{equation}\label{12}
\overline{\cos\phi}=\overline{\cos^2(\phi/2)}-\overline{\sin^2(\phi/2)}=1/2-1/2=0
\end{equation}
The pure (bra+ket) pairs create a time-independent background
while the mixed (bra+ket) pairs create fluctuations over this
background.
\par
Because of this it is reasonable to take the set of pure pairs
(bra+ket) with their occupancy numbers as the initial state of a
quantum system. A perturbation can produce mixed pairs from
initial pure pairs, e.g. the pure pairs $\psi^\dag_p\psi_p$ and
$\psi^\dag_{p'}\psi_{p'}$ can become the mixed pairs
$\psi^\dag_{p'}\psi_p$ and $\psi^\dag_p\psi_{p'}$. These mixed
pairs arise also after the bra or ket exchange between two
occupied states $F_p$ and $F_{p'}$ without any perturbation. Note
that two mixed pairs created by the exchange are phase correlated
since they have equal and sign-opposite phases. The observed
quantum particles represented by such pairs are just the
mysterious entangled quantum particles which are so popular
nowadays. Their magic property is simply the phase correlation.
They have no mysterious links over the entire Universe though
indeed their mutual phase correlation may hold rather long during
their unperturbed evolution.
\par
To avoid any misunderstanding let us emphasize that any action on
one of such phase correlated particles cannot in any way influence
its correlated partner.
\par
Now as an example let us consider the plane wave states with the
momenta $p$ and $p'$. We have for the position operator $U(x)$:
\begin{equation}\label{13}
U(x,t)=F_p+F_{p'}+2\sqrt{F_{p'}F_p}
\cos[(p-p')x+(\epsilon_{p'}-\epsilon_p)t+(\varphi_p-\varphi_{p'})]
\end{equation}
This expression describes the constant background and the time and
space dependent fluctuations. The fluctuation arises when the bra
of one state meets the ket of other state.
\par
We can rewrite (\ref{13}) as a traveling fluctuation wave over the
background:
\begin{equation}\label{14}
U(x,t)/2=\frac{F_p+F_{p'}}{2}+\sqrt{F_{p'}F_p} \cos(qx-\omega
t+\phi)
\end{equation}
Here $q=p-p'$ is the wave vector and
$\omega=\epsilon_p-\epsilon_{p'}$ is the wave frequency and
$\phi=\varphi_p-\varphi_{p'}$ is the wave initial phase.
\par
For small values of the momentum differences taken as
$q\rightarrow\hbar q$ with $\hbar\rightarrow 0$ we get
\begin{equation}\label{15}
 U(x,t)/2\simeq F_p[1+\cos(qx-qvt)]
\end{equation}
Here $v=\partial\epsilon/\partial p$ and $\omega\simeq qv$. Now
the wave vector $q$ and the momentum $p$ become independent
quantities. Summing (\ref{15}) over $q$ we get from it the
classical trajectory of the bra+ket pair with momentum $p$ and the
velocity $v$. Restoring the vector indices we can write the
probability to find this pair as:
\begin{equation}\label{16}
W({\bf R},t)=\sum_q \cos[{\bf qR}-{\bf qv}t)]=\delta({\bf R}-{\bf
v}t)
\end{equation}
This expression show that the so-called quantum particles that we
see e.g. in the Wilson camera or in a photo-plate are really {\it
classical particles}. They are the (bra+ket) pairs moving together
and therefore visible. Such pairs are described by classical
distribution functions or by Wigner functions constructed from the
bra and ket functions.
\par
There are two languages in QM, the wave language and the
corpuscular language. They should be equivalent so we can realize
the bra and ket equally either as waves or as corpuscles. A wave
has a phase, so it is necessary to ascribe phases also to these
bra or ket corpuscles. One may imagine them as the "messengers
with clocks"\, used in [\cite{R1, R2}] for the numerical
simulation of quantum phenomena. Actually are these bra or ket
waves or corpuscles is the detail of secondary importance. It is
crucial only that {\it two quantum entities are necessary to get
an observable quantity in our Classical World}. Note also that two
independent quantities of Quantum World (i.e. the bra and ket)
correspond just to two independent quantities of Classical World
namely an observed quantity and its time derivative.
\par
Since the bra and ket taken alone are undetectable (invisible) in
our world {\it they do not belong to it}. Only their encounter
(described by corresponding QM expressions) makes them detectable
(visible) and corresponds to the measurement or, better say, to
their appearance in our world as {\it classical objects}.
\par
Let us repeat once more that the actual Quantum World is the
invisible world of separate bra and ket. They move independently
and reveal themselves only after the mutual encounter. Our
Classical World is the world of pairs (bra+ket) in the same state
or in the close states moving together and looking as classical
objects. This is the right physical picture for QM which follows
logically from its mathematics.
\par
Suppose that we see a point on a screen or photo-plate which
appeared during an experiment with a quantum particle. This point
definitely belongs to our classical world. In classical world this
point was the point in its previous life as a point-like classical
particle. However, there are no reasons to think that it is so in
the quantum world. We see the point after the measurement and can
say nothing what was before. To say that the point was also the
point before the measurement represents (according to QM
principles!) an illicit extrapolation based on nothing. Of course,
this extrapolation was so natural and self-obvious that it was
accepted explicitly or implicitly by the creators of QM and by
their followers. And despite the fact that the QM mathematics
describing the experiment contains two separate entities, one bra
and one ket with their own space and time coordinates! According
to QM rules we cannot detect them separately but only jointly when
they enter in the detector device in order to appear in our
classical world. The adequate physical picture of quantum
phenomena should necessary be based on the pair (bra+ket moving
together) in our classical world and on the pair (bra and ket as
separate entities) in the quantum world.
\par
In the bra-ket picture the usual QM puzzles and mysteries
evaporated. Of course, instead of the present so popular
"intriguing and fascinating features of quantum world"\, many new
unclear questions will inevitably arise but they will not have
such impasse character. A picture which is fundamentally wrong
inevitably leads to the impasse with no outcome. On the contrary a
basically right picture can be improved and ameliorated rather
easily.
\par
The bra-ket physical picture has a classical analogy. In a gas of
classical particles with pair collisions there is a phenomenon
known as the "long tails" of the response. This long range and
long living correlation is created by the repeated collisions
between two classical particles. The first collision creates the
correlation while the second collision takes it into account. In a
wave language the correlation is described by pairs of correlated
diffusion modes which represent the correlated pairs of gas
particles [\cite{AW,On,GKK}]. The gas remains spatially homogenous
and only pairs of diffusion modes are observable on a large time
and space scale. The kinetics of this phenomenon has many
analogies with QM.
\par
One says usually that a quantum object being not a particle and
not a wave is a "wavicle"\, totally inaccessible to our human
imagination. We see that the mysterious "wavicle"\, may be
reasonably interpreted as a pair of two objects. The QM
expressions for observable quantities look exactly as the
correlation formulae for just two independent variables (with some
weight). Therefore, one may regard quantum mechanics as a kind of
correlation statistics of random events [\cite{Bp}]. The picture
of bra-ket pairs is in agreement with such point of view provided
that the random events are the encounters of these bra and ket and
the random quantities are their phases. The phases may be regarded
as the hidden variables of Quantum Mechanics.
\par
We saw above how the so-called measurement problem vanished. Also
the great mystery of wavicle self-interference reduces to
triviality. A point cannot be in one time in two places but a bra
corpuscle and a ket corpuscle (i.e. the wavicle) can do it quite
easily. They can go through two slits or through only one slit.
The knowledge that the wavicle went through one slit means simply
that its bra and ket were there and therefore the interference is
absent by definition. For two wavicles one can detect them in both
slits and nevertheless observe the interference due to the
possible bra-bra or ket-ket exchange between two wavicles. This
exchange mechanism is also the origin of the mysterious quantum
correlation at large distances known as the EPR-paradox and the
violation of the Bell inequality. Using the bra-ket picture one
can easily explain it [\cite{GG}] in a local way by the "common
cause in the past"\, as it should be in all one-time correlation
phenomena.
\par
In this picture it is possible also to get simple and reasonable
answers on a number of other questions including those that "one
cannot asks".


\begin{thebibliography}{99}
\bibitem{Zel} A.Zeilinger, Rev.Mod.Phys., {\bf 71}, S288,(1999).
\bibitem{Omn} R.Omnes, Rev.Mod.Phys., {\bf 64}, 339, (2002).
\bibitem{Zr} W.H.Zurek, Rev.Mod.Phys., {\bf 75}, 715-775,(2003).
\bibitem{Schl} M.Schlosshauser, Rev.Mod.Phys., {\bf 76}, 1268-1303, (2004).
\bibitem{Prs} A.Peres, Rev.Mod.Phys., 76, 93, (2004)
\bibitem{B1} N.Bohr, "Atomic Theory and the Description of Nature",
London, (1934).
\bibitem{B2} M.Born, "Atomic Physics", London-Glasgow, (1963).
\bibitem{BH} W.Heisenberg, in the essays "Niels Bohr and the development of Physics" ,
 ed.W.Pauli, L.Rosenfeld, V.Weisscopf, London, Pergamon Press LTD, (1955)
\bibitem{LB} L.de Broglie, "L'Electron Magnetique", Paris, Herman ed.,(1934).
\bibitem{Schr} E.Schroedinger, Proc.Cambridge Phil. Soc. 31, 555, (1935).
\bibitem{Bm} D.Bohm, "Quantum Theory" (1956).
\bibitem{R1} H. De Raedt and K. Michielsen, Ann. Phys.(Berlin) 524, 393 – 410 (2012)..
\bibitem{R2} H. De Raedt, K. De Raedt, K. Michielsen, K. Keimpema, and S. Miyashita,
J.Comp.Theor.Nanosci. 4, 957 – 991 (2007).
\bibitem{AW} B.Alder, T.Wainwright, Phys.Rev.Lett., {\bf 18}, 968, (1967).
\bibitem{On} A.Onuki, J.Stat.Phys., {\bf 18}, 475, (1978).
\bibitem{GKK} S.V.Gantsevich, V.D.Kagan, R.Katilius, Zh.Eksp.Teor.Fiz.
{\bf 84}, 2082 (1983), [Sov.Phys.JETP {\bf 57}, 1212 (1983)].
\bibitem{Bp} F.Bopp, Zs.Naturforsch.,{\bf 2a}, 202, (1947);{\bf 7a}, 82, (1952);{\bf 8a}, 6, (1953).
\bibitem{GG} S.V.Gantsevich, V.L.Gurevich, arXiv [quant-ph], 1512.03672, (2015).
\end{thebibliography}
\end{document}